# The Advanced Light Source Accelerator Control System at Ten Years from Commissioning[*]

A. Biocca, W. Brown, E. Domning, K. Fowler, S. Jacobson, J. McDonald, P. Molinari, A. Robb, L. Shalz, J. Spring, C. Timossi, LBNL, Berkeley, CA 94720, USA

*Abstract*

The Advanced Light Source [1] was commissioned 10 years ago using the newly constructed control system [2]. Further experience with the control system was reported in 1993 [3]. In this publication, we report on recent experience with the operation and especially growth of the computer control system and expansion to accommodate the new superconducting bend magnets and fast orbit feedback for the ALS electron storage ring.

## BACKGROUND

The Advanced Light Source is a third generation 2 GEV Synchrotron radiation source with ports for 60 beamlines. ALS was commissioned in 1990 with the control system reported in [2]. This system by S. Magyary et al was based on nearly 500 custom made Intel 80186 based microprocessor controllers (Intelligent Local Controllers or ILCs), connected via a bidirectional optical fiber interconnect carrying modified SDLC to an Intel Multibus based central memory architecture (Collector Multi-Module CMM/Display Multi-Module DMM) with dedicated links to 50 Mhz 486 PC's running Windows 3.11. Accelerator Controls communication with experimental beamlines and endstations was via EPICS CA (Experimental Physics and Industrial Control System Channel Access [4]) to process variables instantiated in a multiprocessor VME crate interfaced via Bit-3 bus couplers to the Multibus shared-memory core. VME based EPICS systems were also deployed in the storage ring for vacuum readouts and the early beamline controls were also EPICS based. The initial networks were 10 megabit thicknet wiring to a Cisco 7500 series router with a 100 megabit FDDI link to the LBNL site central network.

### Software and Data Flow

Computer control of the ALS is distributed amongst embedded processors deployed near the hardware connected remotely to display computers in the control room.

Control room supervisory and status applications run on a dozen PCs, collectively known as 'console computers' and are developed with a plethora of languages including Microsoft Visual C and Basic, Borland's Delphi and National Instruments' Labview. More active control programs, such as orbit feedback, are developed primarily with MathWorks' Matlab and are deployed on Sparcstations running Solaris.

Programs and applications for hardware control run on embedded processors in one of two configurations: Programs running on the ILC are developed in C or PL/M and are downloaded over the serial network. A configuration database is also loaded to describe the devices to be controlled. The VME and Compact PCI (cPCI) subsystems run EPICS on VxWorks and applications are a combination of real-time databases and C code.

Data and control flows between the console computers and the process controllers follow two distinct paths. The original path traverses a massively parallel serial RS485 and optical fiber network using a simple SDLC based protocol. Data is polled continuously to the central memory (CMM) from the 500 ILCs over 50 multi-drop links. From the central memory, data is polled on demand to the console computers across another set of dedicated serial links. As new accelerator systems are added, and for older systems needing higher performance, the original data path is replaced by a dedicated Ethernet subnet using the EPICS Channel Access protocol.

These different data paths present a challenge for software development on the console computers. Ideally, the differences should be transparent, and significant work has been done to achieve this goal. A dedicated multiprocessor EPICS based Input/Output Controller (IOCs) is bus-coupled to the central memory making the network of ILC data available via the Ethernet. Many applications for the console computers have been developed using an interface library, deployed as a windows Dynamic Link Library

---

[*] Work supported by the Director, Office of Energy Research, Office of Basic Energy Sciences, Material Sciences Division, U.S. Department of Energy, under Contract No. DE-AC03-76SF00098

(DLL), with an API designed for the original serial network. Changing and re-testing all these programs to use the channel access API is not feasible due to the large effort required. Instead a compatibility library has been deployed that mimics the behavior of the original library. Since linking is dynamic, applications don't have to be rebuilt.

Both the compatibility library and the newly developed applications use a C language API called Simple Channel Access (SCA), which was designed to make EPICS Channel Access client development less complex. SCA is now widely used throughout the facility, both on Windows and Unix, and has been linked with software development packages such as Matlab and Labview.

## CURRENT CONFIGURATION

The Console Computers have been upgraded several times, first to Windows NT 3.51, and later to NT 4.0, and they are now in the process of being migrated to Windows 2000 Professional. The hardware has also been updated several times and now all console PC's are at least 400 MHZ Pentiums. Additionally, three Sun Workstations have been added to the console to handle EPICS based operator interfaces for the Damping Systems and to run Matlab based control software. Matlab has become the high-level programming environment for ALS control and many of the accelerator physics experiments.

### Servers

PC and Unix servers have been upgraded several times, from the original Digital 486 and Sparc 2/IPX's to more current PCs and Sun Netra's. The more significant change in the server arena was the consolidation of file services into Network Attached Storage. A Network Appliance 720 performs this duty with 300GB of FibreChannel disks and a 1 Gig network attachment. Backups are to a locally attached SCSI DLT library. This serves both NFS and CIFS clients and runs no user code so availability is extremely high. Rebooting servers no longer takes filesystems down.

### Equipment Protection System

Equipment Protection Systems (EPS) at the Light Source are implemented using Modicon PLC type equipment. A Modbus serial interconnect has been implemented to EPICS VME crates and a read-only program polls selected PLC memory locations into EPICS Process Variables, making them available to the system and Data Archiving. This facilitates documentation of beam time usage and aids in fault analysis.

Data Archiving was implemented initially using an EPICS diagnostic tool that was not designed for continuous production operation. A more reliable Archiving engine was implemented based on our local SCA library that has proven itself in several years of use. Web and application program interfaces allow data to be mined for many purposes.

### Hardware Controllers

Power Supplies and Instrumentation interfacing in the original controls were implemented via the custom manufactured ILCs. They continue to be used, but in applications requiring greater performance they are being replaced with a Compact PCI based solution. Compact PCI was chosen after a careful evaluation of all requirements including the compatibility required with the installed facility racking and cabling. Compact PCI offers significant rear I/O and adequate capability configured in a 3U rack package, matching the existing equipment racking and cabling. COTS (Commercial Off the Shelf) hardware was employed as much as possible, but to meet requirements a slightly customized commercial enclosure coupled with a custom rear I/O PC board, and a custom Industry Pack (IP) board allowed us to meet ALS requirements with excellent performance at reasonable cost. The IP board carries four 16 bit analog input channels, two 16 bit analog outputs, and two bytes of binary I/O. A pair of these IP cards exceed the I/O capability of one ILC, and four will fit on each IP carrier board. They were designed in-house and assembled by outside vendors. The cPCI CPU selected is a 300 Mhz PowerPC board from Motorola (MCP750) running EPICS on VxWorks. The IP Carrier is from SBS GreenSpring. Fifteen of these chassis are deployed, enough to handle all of the power supplies in the Storage Ring. Presently only the correctors and the main strings are under cPCI control and readouts of some of the beam position monitors, the rest will be moved as needed. Twelve of the cPCI chassis were installed with their own private 100 Megabit network subnet and a fast orbit feedback system with a 1 khz update rate is in development, with beam position information sharing to take place via multicasts over this network [5].

### Facility Networks

The original network was based on a Cisco 7509 router. A separate subnet was employed for each beamline sector area, and one for the accelerator controls. This has been upgraded to a Cisco 8530 primary router with 40 Gigabit backplane and 1 Gig connections, and a 7513 router was configured as a redundant backup. Managed switches are deployed at each facility sector, and several in the accelerator controls subnets. Unmanaged switches are deployed only to non operationally critical subsystems and hosts. The number of accelerator controls subnets was increased to three, one each for controls and services plus a separate one for the fast orbit feedback controls. Additional access security is provided by configuring the routers to limit connectivity of the control networks to local subnets. The legacy thicknet network cabling is in the process of being decommissioned, and all ports are being converted to Category 5 10/100 Base-T or better. Wireless 802.11B networking has been installed and is used for debugging from laptops.

*Devicenet*

The recent replacement of three ALS bend magnets with superconducting units necessitated 18 bit control precision on their power supplies. To achieve this and minimize noise, etc., it was decided they be Digitally controlled, and DeviceNet was selected as the interface. After a failure to get the SBS IP DeviceNet card to work properly we changed to an SST 5136 VME board for production operation. Labview DeviceNet boards were used for testing and development.

## CORE UPGRADE

The Controls Core Upgrade is currently in progress. The goal is to retire the difficult to support core components of the control system. These have been identified to include the Multibus-1 and –2, the PC SDLC links and the bi-directional fiber link equipment. The methodology is to develop an EPICS Ethernet to ILC link adapter (hardware, software, and compatibility libraries), and then deploy these throughout the facility, connecting to the copper ILC links already in place. The adapter under development now consists of a VME CPU and an SDLC capable multifunction serial I/O board on IP (both commercial products). The software to communicate with the ILCs is in development now. This project will also require new Operator Knob Panels, and a prototype has been developed based on optical encoders, LCD displays, and an Atmel AVR microprocessor.

Current network upgrades include increasing the number of 1 Gig links to hosts and retiring all remaining thicknet network cabling.

The corrector magnet power supplies are presently controlled with 16 bit DACs, but slightly smaller step size is needed to minimize beam motion during global orbit feedback. A new rear I/O PC board is under test that adds a second DAC to each control for finer resolution. This fine DAC is to be controlled by the feedback algorithm, while the setpoint is placed in the coarse DAC. A prototype has been constructed and software development for testing is underway.

## FUTURE IS NETWORKING

Clearly the Network will continue to be upgraded as beamlines have increasing bandwidth requirements, and the core will likely go to 10 Gigs and beyond. Host connections of 1 Gig will become commonplace, and the performance and reliability of network equipment will improve.

One industry trend in I/O is toward the use of direct Ethernet in the whole price range of devices. When we were selecting DeviceNet for our SuperBend controls one goal we had was to choose a field bus that would facilitate low cost I/O. At that time Ethernet did not meet the cost requirement. Now Ethernet devices are under $100 and an Ethernet protocol stack will fit in a $7 embedded CPU. This low cost makes it possible to build a distributed control and data acquisition system using off-the-shelf modules that have Ethernet interfaces, placing them near the controlled equipment to minimize cabling, and use standard networking to tie the system together. In essence the whole system becomes a high performance network coupling various types of computing and I/O resources together. The bandwidth of this network is sufficient to do things that previously required special and costly single-vendor interconnects. As links between processors reach 1G and TCP/IP QOS (quality of service) become mature it will be possible to meet more system requirements with standard networking, at lower initial and life cycle cost.